\documentclass[aps,showpacs,twocolumn,twoside]{revtex4}

\usepackage{epsfig}
\usepackage{amsmath,amssymb,color}

\usepackage[english]{babel}

\parskip=\medskipamount

\newcommand{\eq}[1]{(\ref{#1})}
\newcommand{\fig}[1]{Fig.\ref{#1}}

\newcommand{\be}{\begin{equation}}
\newcommand{\ee}{\end{equation}}
\newcommand{\barr}{\begin{array}}
\newcommand{\earr}{\end{array}}
\newcommand{\beqn}{\begin{eqnarray}}
\newcommand{\eeqn}{\end{eqnarray}}
\newcommand{\bs}{\begin{subequations}}
\newcommand{\es}{\end{subequations}}

\begin{document}

\title{Fractal globule as an artificial molecular machine}

\author{V.A. Avetisov$^{1,5}$, V.A. Ivanov$^2$, D.A. Meshkov$^1$, S.K. Nechaev$^{3,4}$}

\affiliation{$^1$N.N. Semenov Institute of Chemical Physics, RAS, 119991 Moscow, Russia \\
$^2$Faculty of Physics of the M. V. Lomonosov Moscow State University, 119991 Moscow, Russia \\
$^3$Universit\'e Paris-Sud/CNRS, LPTMS, UMR8626, B\^at. 100, 91405 Orsay, France \\ $^4$P.N.
Lebedev Physical Institute, RAS, 119991 Moscow, Russia \\ $^5$ National Research University Higher
School of Economics, 109028 Moscow, Russia}

%\date{\today}

\begin{abstract}

The relaxation of an elastic network, constructed by a contact map of a fractal (crumpled) polymer
globule is investigated. We found that: i) the slowest mode of the network is separated from the
rest of the spectrum by a wide gap, and ii) the network quickly relaxes to a low--dimensional
(one-dimensional, in our demonstration) manifold spanned by slowest degrees of freedom with a large
basin of attraction, and then slowly approaches the equilibrium not escaping this manifold. By
these dynamic properties, the fractal globule elastic network is similar to real biological
molecular machines, like myosin. We have demonstrated that unfolding of a fractal globule can be
described as a cascade of equilibrium phase transitions in a hierarchical system. Unfolding
manifests itself in a sequential loss of stability of hierarchical levels with the temperature
change.

\end{abstract}

\pacs{61.41.+e, 64.60.al, 64.60.aq, 87.15.Cc}

\maketitle

A notion of a molecular machine (MM) is usually attributed to a nanoscale structure able to convert
perturbations of fast degrees of freedom into a slow motion along a specific path on a
low--dimensional manifold -- see \cite{Blum,Blum-Gros,Mikh}. Rich diversity of MMs is produced in a
living world. Molecular machines possess various functions: they create or break chemical bonds,
move mesoscopic objects, etc. Globular proteins formed by polypeptides of hundreds units look as
simplest examples of MM. The ability of proteins to manipulate accurately by solitary charges, or
atoms against fluctuations, is the main requirement for reproduction of living systems at the
molecular level \cite{Eigen, Avet-Gold}.

In the definition of MMs we follow the work \cite{Mikh}, where a MM is specified by a particular
relaxation dynamics of its elastic network. Represent a molecular structure by a spatial
distribution of a set of nodes $i=1,...,M$ and undirected links between them, encoded in an
adjacency matrix ${\bf A}$ with the elements $a_{ij}=1$ for a link between $i$ and $j$, and
$a_{ij}=0$ otherwise. In elastic networks, linked nodes are subjected the action of elastic forces
that obey the Hooke's law. In the over-damped limit, the velocity of each node is proportional to
the sum of applied elastic forces, so the network relaxation can be written as:
\be
\begin{array}{lll}
\dot{{\bf R}}_i(t) & = & \sum_{j=1}^M a_{ij} {\bf u}_{ij} \left(|{\bf R}_i-{\bf
R}_j|-|{\bf R}_i^{(0)}-{\bf R}_j^{(0)}|\right) \medskip \\
\dot{{\bf r}}_i(t) & = & -\sum_{j=1}^M{\bf \Lambda}_{ij}{\bf r}_j
\end{array}
\label{eq:1}
\ee
where ${\bf R}_i(t)\equiv {\bf R}_i$ and ${\bf R}_i^{(0)}$ are the current and equilibrium
positions of node $i$, ${\bf u}_{ij}=\frac{{\bf R}_i-{\bf R}_j}{\left|{\bf R}_i-{\bf R}_j\right|}$,
and ${\bf r}_i={\bf R}_i-{\bf R}_i^{0}$ are the small linearized deviations from equilibrium. The
strain tensors, ${\bf \Lambda}_{ij}$, are the building blocks of the $3M\times 3M$ linearized
matrix ${\bf \Lambda}$. The relaxation of a system under small perturbations is a sum of
independent normal modes, $|{\bf r}_i(t)|=\sum_{k=1}^{3M}{\bf r}_i(0)\,{\bf e}_k e^{-\lambda_k t}$,
where $\lambda_k > 0$ and ${\bf e}_k$ are the eigenvalues and eigenvectors of ${\bf \Lambda}$.

The distinguished features of MMs became transparent by investigating the relaxation properties of
protein elastic networks like myosin \cite{Mikh}. Being perturbed, the elastic network quickly
reaches a low-dimensional attracting manifold spanned by slowest degrees of freedom, and then
slowly relaxes to the equilibrium along a particular path in this manifold. The existence of a wide
spectral gap separating slowest (soft) and fast (rigid) modes, and a low-dimensional manifold with
a large basin of attraction, define a generic molecular machine.

Is it possible to design molecular machines artificially? This challenging question addresses to
the prospects of artificial algorithmic chemistry. It concerns the problem of overcoming the
``complexity threshold'' in prebiology, meaning that even a ``primary'' cell with a minimal
machinery of biological replication remains hopelessly complex for its spontaneous appearance
beyond the world of MMs (see, for example, \cite{Koonin}). So, the question of how the ``primary''
MMs occurs under prebiotic conditions is open. In this paper the polymer globules with properties
of MMs are the targets of our search.

It is well-known that the non-phantomness of a polymer chain in $3D$ space leads to: i) bulk
interactions vanishing for infinitely thin chains, and ii) topological constraints, preserved even
for zero's thickness chains. At $T>\theta$, a polymer of $N$ segments, each of length $a$, is in a
coil phase, while at $T<\theta$, it collapses into a weakly fluctuating drop-like globule of size
$R_{gl}\sim a N^{1/3}$ (\cite{Lifsh, Gros-Khokh}). In the globular phase formed by a linear chain
\emph{with open ends}, all subchains of length $a s$, for $s \lesssim N^{2/3}$, look as Gaussian
coils since the volume interactions are screening in the melt \cite{deGennes}. Such globules have
almost random elastic networks with many local metastable states and small basins of attraction.
The soft and rigid relaxation modes of random networks are poorly separated \cite{Mikh} meaning
that ordinary equilibrium polymer globules do not possess MM properties.

\begin{figure}[ht]
\epsfig{file=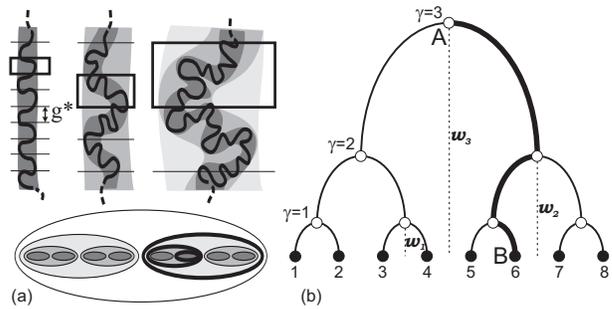,width=8cm}
\caption{a) The hierarchically folded part of a polymer chain is identified with hierarchically
embedded basins (crumples); b) The boundary of a Cayley tree is the space of states for
hierarchical system. The path AB corresponds to a specific sequence of crumples in (a).}
\label{fig:01}
\end{figure}

For linear \emph{initially unknotted} polymer rings the situation is quite different
\cite{Gros-Nech-Shakh}. The chain folds into a weakly entangled on all scales, space-filling
equilibrium crumpled (fractal) globule (FG). The FG structure can be elucidated by the following
imaginative recursive description. At $T<\theta$ there exists a certain length, $g^*\simeq
N_e/(a^6\rho^2)$, such that chain parts larger than $g^*$ collapse ($N_e$ is the well-known
parameter of the reptation model \cite{doi} and $\rho$ is the globule density). Define $g^*$ as the
block monomers, constituting the crumple of a minimal (1st-level) scale. The chain segments
containing several consecutive 1st-level crumples collapse ``in its own volume'', forming the
2nd-level crumples, etc. The chain packing is completed when all $g^*$-length folds belong to a
single crumple of the largest level. Such a process suggests a fractal dimension $D_{f}=D=3$ (space
filling) -- see the \fig{fig:01}a. Note that our recursive construction pursues mainly illustrative
aims, being a convenient way to describe the final equilibrium globular conformation, where any two
crumples larger than $g^*$, which are neighboring in space, are weakly mutually entangled on all
scales. The computer simulations of a linear chain collapse after a temperature jump show
\cite{kinetics} that the spontaneously formed 1-st level crumples experience instability and the
larger crumple swallows up the chain material from smaller ones until the whole globule is formed.
Meanwhile, extensive numeric computations of Alexander invariants of equilibrium  subchain
conformations in a 108~000-monomer collapsed unknotted ring \cite{mirny-im-nech} demonstrate that
the final structure is weakly knotted in a broad interval of scales and crumples do separate
topologically. This makes the concept of hierarchical crumpling used in \cite{rost} for
self-consistent description of collapse and swelling dynamics (but without account for topological
effects) plausible.

In a FG each $g^*$-length unit is characterized by a set of indices specifying to which particular
1st-level fold (embedded into 2nd-level fold, \emph{etc}) this unit belongs. Such indexing is
encoded by a terminal leaves (black dots) of a Cayley tree, defining a unique path (AB) from the
tree root to the boundary, thus fixing the unique hierarchy of folds -- see the \fig{fig:01}b. The
distance along a tree from a leave to the root point, measured in the number of generations,
designates the current hierarchical level. The terminal leaves constitute a space of states for the
units, and each subtree is in one-to-one correspondence with a particular crumple. The natural
parametrization of a hierarchically folded globule is the special ``system of indices'' enumerating
the terminal leaves of a Cayley tree, known as ``$p$-adic numbers'' \cite{VVZ}.

Thus, the conformations of hierarchically folded chain are in bijection with the trajectories of a
walker in an \emph{ultrametric space}. The ultrametric distance between two distinct terminal
leaves sets a special (non-Archimedian) metric at the Cayley tree boundary: the transition between
two distinct terminal leaves is determined by the highest barrier (largest crumple) separating
them. This concept can be straightforwardly translated into the space of hierarchically folded
configurations: as shorter the distance between two leaves at the tree boundary, as easier the
passage from one folded configuration to the other one. This is reminiscent to the description of
topological states of a path entangled with a regular array of obstacles. In the latter case the
chain entanglement in the lattice with the coordinational number $p$ is fully characterized by the
configuration of the shortest (primitive) path along a $p+1$--branching Cayley tree
\cite{khokh,sem}. The Cayley tree serves as an expanded (``covering'') space, which simultaneously
defines the coordinates of the chain ends, and the chain topology \cite{nech}: as shorter the
primitive path, as less entangled the chain. Appealing to the described analogy, one sees also the
difference between entanglements and crumpling. For entanglements the metric is fixed by the
distance \emph{along} the Cayley tree, while for crumpling -- \emph{at the boundary} of the tree.

Let $f_{i}(s)$ be the probability to find a walker in the state $i$ at the boundary of the tree
after $s$ steps ($0\le s \le N$), if at $s=0$ it was at $i_0$. The transition function $f_i(s=N)$
defines a measure of hierarchically folded chains of lengths $N$.  For example, $f_{i=0}(s=N)$ is a
statistical weight of all hierarchically folded conformations of a chain of length $aN$ whose ends
are in one and the same minimal fold. Since the random walk is a homogeneous Markov process, the
transition function $f_i(s)$ satisfies the Kolmogorov--Feller equation \eq{eq:3}
\be
\frac{\partial f_i(s)}{\partial s}= \sum_{j\neq i}w(i|j) f_j(s) - \Big[\sum_{j\neq i}w(j|i)\Big]
f_i(s)
\label{eq:3}
\ee
subject to the initial condition $f_i(s=0)=\delta_{i,i_0}$, where $w(i|j)=w(j|i)$ is the
probability to jump between $i$ and $j$ in one step -- see \cite{Stein,Avetisov1,Avetisov2} and
Supplement 1.

Unfolding at some scale, $\gamma_{cr}$, means the destruction of a hierarchical tree--like set of
states above $\gamma_{cr}$. This can be easily achieved by introducing an auxiliary
temperature-dependent repulsion between the units. The truncation of these interactions depends on
the scale of crumples. To satisfy this requirement, the interaction potential should be again of a
block-hierarchical form ${\bf U}$ with elements depending on the ultrametric distances between the
units only. We can generalize \eq{eq:3} to:
\be
\frac{\partial f_i(s)}{\partial s} =  \sum_{j\neq i}w(i|j) f_j(s) - \Big[\sum_{j\neq i}w(i|j)\Big]
f_i(s) + \tau {\bf U} f_i(s)
\label{eq:5}
\ee
where $\tau=\frac{T-\theta}{\theta}$ and ${\bf U}f_i(s) = \sum_{j\neq i}w(i|j)f_j(s)$. Solving
\eq{eq:5}, we get the following nonzero eigenvalues:
\be
\lambda_{\gamma}= -p^{-\alpha\gamma} + \tau\left(-p^{-{\alpha\gamma}} +\frac{1 - p^{-1}}{p^{\alpha}
- p^{-1}}\, p^{-\alpha}\right)
\label{eq:6}
\ee
The solutions of the equations $\lambda_{\gamma}(T)=0$ define a sequence of critical temperatures
$T_1< T_2<... <T_{\rm max}$ at which the hierarchically folded conformations lose the stability. At
$T_1$ the largest fold, i.e. the highest level $\gamma_{\rm max}$, becomes unstable and the fractal
globule unfolds into a set of smaller crumples. At $T_2>T_1$, these crumples of level
$\gamma=\gamma_{\rm max}-1$ unfold as well, \emph{etc}, until $T_{\rm max}$ is reached, at which
the crumple $\gamma=1$ becomes unstable and the FG melts completely.

Thus, the fractal globule is characterized by a set of order parameters, encoding the crumples of
different hierarchical levels. These order parameters have natural representation in terms of
coordinates in the ultrametric space. Unfolding of a fractal (crumpled) polymer globule is
described as a cascade of equilibrium phase transitions in a hierarchical system and manifests
itself in a sequential loss of stability of hierarchical levels with the temperature change.

The apparent simplicity of a FG creation (the abrupt change of the solvent quality provokes a
collapse transition in a long unentangled polymer coil), has led to regard FG as a strong candidate
for a molecular machine. Below we provide numeric simulations demonstrating that the elastic
network (EN) of the fractal globule under some conditions possess the key properties of molecular
machines, i.e. it has the wide spectral gap and the low-dimensional relaxation manifold.

Crumpling of a single linear chain has been simulated by means of the Monte Carlo method and the
bead-spring model in 3D continuous space \cite{monte}. Two different methods of FG creation have
been used. Firstly, the FG has been created by an artificial hierarchical potential applied to a
polymer chain. This protocol allows the precise control of folds structure and formation of
corresponding clusters of links. Secondly, FG is obtained by a fast collapse of a polymer chain
with open ends in a spherically symmetric linear potential. This offers a principle possibility of
an experimental synthetic FG design with the properties of molecular machines.

\emph{Equilibrium collapse in an hierarchical potential}. The polymer chain consisting of $N$ beads
(monomeric units) is modelled in a standard set of interaction potentials \cite{monte}, accounting
for rigidity and volume interactions. In addition, $M$ units are selected (randomly along a chain)
to be specific ones and a hierarchical potential $U_{\gamma}\sim \gamma^{-1}$ operating between the
specific units at the hierarchical level $\gamma$ is introduced. It ensures that on each level
$\gamma$, the formation of clusters of specific units becomes possible if and only if two clusters
of the level $\gamma-1$ are close to each other in space, and vice versa: any cluster of the level
$\gamma-1$ cannot disappear if it is a part of a cluster of the level $\gamma$. The spatial
distance between two clusters of the level $\gamma$ is determined as the smallest distance between
nodes of these clusters. The rules of clustering are very important. They imitate the hierarchical
constraints imposed on the chain conformations under crumpling.

\begin{figure}[ht]
\epsfig{file=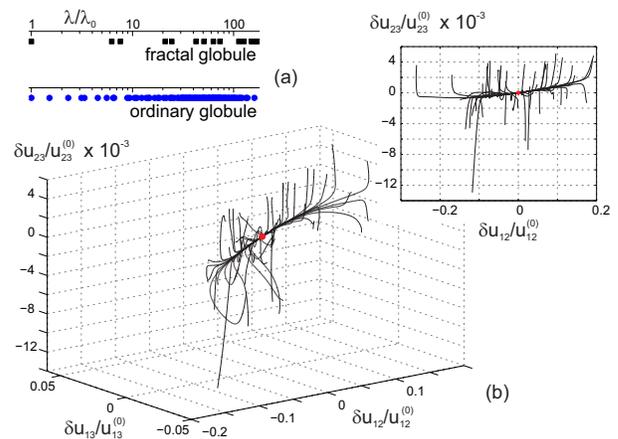,width=8cm}
\caption{(a) Eigenvalue spectra of the network of hierarchically folded chain and ordinary globule
normalized to the lowest nonzero eigenvalue; (b) A set of relaxation trajectories of a
hierarchically folded globule. Insert: Projection of relaxation trajectories on the plane $\delta
u_{23}/u_{23}^{(0)}$ (fast mode) \emph{vs} $\delta u_{12}/u_{12}^{(0)}$ (the slowest mode).}
\label{fig:02}
\end{figure}

Equilibrating the chain by means of Metropolis algorithm at some temperature below the
$\theta$-point, we construct an elastic network: specific units become the nodes of this network
connected by new links if the distances between them are less than the cutoff radius, $R=3$ (in
diameters of a monomeric unit); all other units are washed out. Particular conformation of a
hierarchically collapsed polymer chain and its elastic network are shown in the \fig{fig:sup01}a
and \fig{fig:sup01}b of Supplement 2. The typical eigenvalue spectra of $M$-link elastic networks,
obtained by a hierarchical $N$-monomer chain collapse in a large box without periodical boundary
conditions is shown in the \fig{fig:02}a for $M=32$ and $N=200$. Note a large spectral gap
separating the slowest mode from the rest of spectrum ($\lambda_2/\lambda_1\approx7$,
$\lambda_2/\lambda_3\approx1$). The 3D-view of relaxation trajectories of the hierarchically folded
elastic network (\ref{eq:1}) are plotted in the \fig{fig:02}b.

The hierarchically folded globule quickly relaxes to a low-dimensional (one-dimensional, in our
demonstration) manifold with a large attracting basin, then slowly moves along it to the
equilibrium, highly similar with the myosin dynamics \cite{Mikh}. It should be pointed out that the
elastic network of the fractal globule is highly anisotropic: along the $\delta u_{23}$--axis it is
$10^3$ more rigid that along $\delta u_{12}$-- and $\delta u_{13}$--axes. In an ordinary globule
such strong anisotropy is absent and its elastic network does not have any low--dimensional
attracting manifold for phase trajectories.

\emph{Fast collapse of linear coil in a central linear potential.} In this setting, each monomer of
a linear polymer with open ends experiences the action of an attractive external potential $U_{\rm
ext}$. The potential is switched on during a time $\tau\sim N^2$, where $N$ is the number of
monomers, then it is switched off, the chain configuration is freezed and the network of $M$
uniformly distributed along a chain links is created with the cutoff radius $R$. The theoretical
prediction \cite{Gros-Nech-Shakh} and experiments on kinetics of collapse of polysterol in
cyclohexane \cite{chu}, give $\tau\sim N^2$ for the fast stage of a collapse during which the chain
gyration radius, $R_g$, is stabilized at $R_g\sim N^{1/3}$.

\begin{figure}[ht]
\epsfig{file=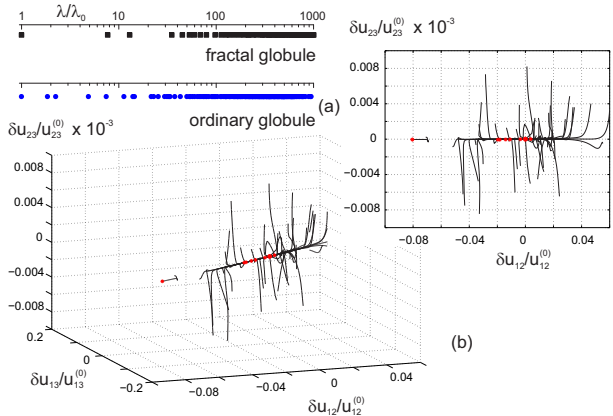,width=8cm}
\caption{The same as in the \fig{fig:02} for a fast chain collapse in a linear radial
potential from a good solvent regime ($\varkappa=0.5,\, R=3,\, M=64,\, N=400$).}
\label{fig:03}
\end{figure}

In simulations we have chosen $U_{\rm ext}({\bf R}_i)=\varkappa |{\bf R}_i|$, where ${\bf R}_i$ is
the $i$th monomer position. The system is placed in a good solvent and the coil-to-globule
transition occurs due to $U_{\rm ext}$ only. The optimal fraction, $M/N$, for which the collapsed
chain typically looks as MM, numerically is found as $M/N\approx 0.16$. In the \fig{fig:03} the
eigenvalue spectrum and the set of relaxation trajectories for $\varkappa=0.5,\, R=3,\, M=64,\,
N=400$ is shown (compare to \fig{fig:02}). Testing different $\varkappa, R, M$ and $N$, we found
that to be MM, the elastic network made by a polymer chain, should: i) not have too many
single-connected vertices; ii) be neither too dense nor very loose.

The hint for choice of parameters follows from a simple mean-field analysis of a phase transition
similar to \cite{Gros-Nech-Shakh}. The polymer chain free energy in a potential $U_{\rm ext}$ can
be estimated as $F = F_{\rm el} + F_{\rm int}$, where $F_{\rm el}$ is the entropic free energy
part, while the interaction part, $F_{\rm int}$, is written in a virial expansion: $F_{\rm int}=
(B\rho^2+C\rho^3)V$, where $B=\frac{T-\theta}{\theta}<0$ and $C$ are the second and third virial
coefficients, and $V$ is the globule volume. The entropic part, $F_{\rm el}$, of an ideal $N$-link
polymer collapsed in $U_{\rm ext}$ in 3D can be estimated by finding the ground state of a
diffusion equation:
\be
\partial_N G(r,N)=D \Delta G(r,N) + \varkappa |r|G(r,N)
\label{eq:diff}
\ee
where $D$ is the diffusion coefficient and the boundary conditions $G(r=0,N)=G(r\to \infty,N)=0$
are assumed. Writing the solution of \eq{eq:diff} in the limit $r\to\infty$ in spherical
coordinates and neglecting the contribution from the gradient term $\partial_r G(r,N)$, we end up
with the Airy equation which has the eigenvalue spectrum $\lambda_{k} = |\zeta_k|D^{1/3}
\varkappa^{2/3}$, where $\zeta_k$ ($k=0,1,...$) denote zeros of the Airy function $Ai(z)$ for
$z=(\lambda-\varkappa r)/(\varkappa^{2/3}D^{1/3})$. The ground state contribution to $F_{\rm el}$
in a linear potential in thermodynamic limit can be estimated as
\be
F_{\rm el} = N \lambda_0 = N |\zeta_0|D^{1/3} \varkappa^{2/3} \quad (\zeta_0\approx -2.338)
\label{eq:airy}
\ee
In \cite{Gros-Nech-Shakh} the entropic part of the FG free energy has been estimated as $F_{\rm el}
= N/g^*$ ($g^*$ is defined above). Equating this expression to \eq{eq:airy}, we get $\varkappa\sim
N_e^{3/2}/(a^9 D^{1/2}\rho^{3})$. Collapsing a chain in a linear potential and keeping this
relation between parameters, one reproduces the main features of a fractal globule formation.

The mechanism which allows to consider FG as a strong candidate for a generic molecular machine
consists in sequential energy transfer from small crumples to larger ones. Few largest crumples
produce deformations on scales of the whole globule. The effect is better visible on highly
asymmetric shapes of polymers, meaning some limitation on collapsed chain length, which may depend
on temperature and polymer rigidity.

The diversity of MM concerns mainly the number of principal soft modes, manifested in the
dimensionality of the attracting manifold. Typically, these manifolds are one- or two-dimensional.
This fact might be important to get functional variability without altering the structural
archetype. The hierarchically folded domain might play in a molecular machine the same constructive
role as an engine in a mechanical device. The possible way of MM design consists, apparently, in
operating with linear polymer systems in external potentials acting in a wide range of scales.
Moreover, the construction of MM is released by preparing them in non-equilibrium conditions. For
example, the pulse irradiation of a linear polymer with topological constraints during a sharp
collapse would allow to produce crosslinks and thus preventing the chain swallowing seen in
\cite{kinetics}. This could result in a formation of a necessary multi-scale structure.

The authors are grateful to A.S. Mikhailov, L. Mirny and M. Imakaev for encouraging discussions.
This work was partially supported by the grant ANR-2011-BS04-013-01 WALKMAT and by MIT--France Seed
Fund. V.A.A. thanks the National Recearch University Higher School of Economics for supporting this
work. Computer simulations have been performed at the Supercomputing Center of MSU
\cite{supercomp}.

\begin{appendix}

\section{Supplement 1}

\begin{figure}[ht]
\epsfig{file=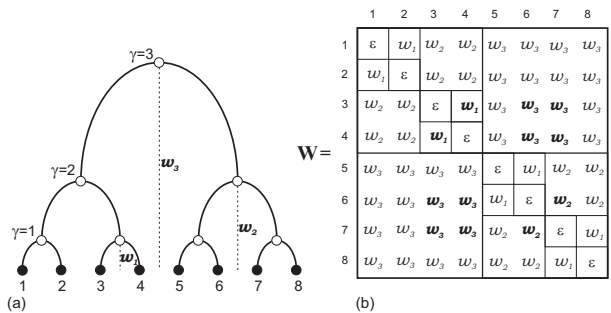, width=8cm}
\caption{a) Space of states in a fractal globule. Each chain monomer is characterized by its position
on the boundary of the ultrametric space (black dots); b) Block--hierarchical structure of
transition matrix ($p=2$).}
\label{fig:sup01}
\end{figure}

The transitions between different states and their correspondence with the Parisi matrix $W$ (see
\cite{Parisi,Mezard} for details) is shown in the \fig{fig:sup01}.

For a random walk at the boundary of a regular $p$--branching Cayley tree, each $\gamma$--level
block of Parisi matrix, ${\bf W}$, consists of $p^\gamma$ equal elements, that depend on $\gamma$
only. For example, $w(3|4)=w_1,\, w(3|7)=w_3,\, w(5|8)=w_2$,  \emph{etc} -- compare
\fig{fig:sup01}a and \fig{fig:sup01}b. The random walk in the ultrametric space is specified by a
choice of transitions $w_{\gamma}$ which are supposed to be $w_{\gamma}=p^{-(\alpha+1) \gamma}$,
where $\alpha>0$ is some parameter -- see \cite{Avetisov1}. The exponential decay of transition
probabilities is the most common situation in nature and, as it is shown in \cite{alpha=1}, is
fully consistent (for $\alpha=1$) with the contact probabilities in a human chromosomes map
provided by HiC method in \cite{Mirny}.

It is known (\cite{VVZ}) that the symmetric transition operator in the right hand side of \eq{eq:3}
is diagonalizible by the $p$-adic Fourier transform, and for unbounded ultrametric space its
non-zero eigenvalues are $\lambda_{\gamma}= -p^{-\alpha\gamma}$. Since all eigenvalues are
negative, the hierarchically folded conformations given by \eq{eq:3} have a well defined ground
state with a specific \emph{set of eigenvalues, one for each fold of a given scale}. The chain
conformations are subjected to the action of a hierarchy of strong constrains, and the temperature
is supposed to be much lower than the critical temperature of the coil-globule phase transition.

It has been shown in \cite{Avet-Nech3} that it is naturally to describe the transitions in a
hierarchical network by the critical behavior in a system of Ising spins with block-hierarchical
interactions, the so-called ``hierarchical Dyson model'' \cite{Dyson,baker}. The Dyson model has a
family of phase transitions with decreasing temperature, producing a tree of hierarchically
embedded clusters of correlated spins. This behavior suggests a new way of analytic approach to a
fractal globule unfolding, based not on the coordinate representation, which defines the position
of a given monomer in a real Euclidean space, but on a natural ``intrinsic'' $p$-adic encoding of
folds, which attributes a given unit to a particular set of crumples of all hierarchical levels.
The precision of such a description is restricted by the chain length in the unit, i.e. in the
crumple of minimal scale, $g^*$. Such a description \emph{is not a theory of coil--to--fractal
globule} phase transition in a common sense, but rather the \emph{theory of a fractal globule
melting}, since we are interested in the destruction of hierarchical set of folds only (without
refereing to their coordinate representation), which happens \emph{within} the globular phase of a
polymer.

\section{Supplement 2}

\begin{figure}[ht]
\epsfig{file=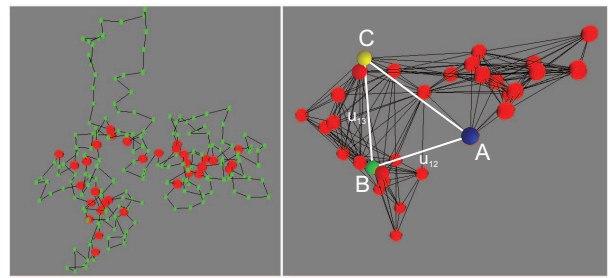,width=8cm}
\caption {(a) Typical shape of hierarchically folded globule. Representative nodes are shown in
red, other monomers are green; (b) The elastic network. Blue (A), green (B) and yellow (C) colors
mark the three nodes constituting the 3D coordinate system for the visualization of the network
dynamics.}
\label{fig:sup02}
\end{figure}

A typical conformation of hierarchically collapsed polymer chain and its elastic network are shown
in \fig{fig:sup02}a and \fig{fig:sup02}b respectively. The benchmarks A, B and C depicted in the
\fig{fig:sup02}b are used to visualize the relaxation trajectories of the elastic network (see
\cite{Mikh} for details).

To represent the slowest degrees of freedom, a triangle ABC (marked by blue, green, and yellow
nodes in the \fig{fig:sup02}b) is chosen in such a way that the changes of the distances along the
sides AB ($\delta u_{12}/u_{12}^{(0)}$) and BC ($\delta u_{13}/u_{13}^{(0)}$) correspond to the
slowest relaxation modes. The particular selection of benchmarks A, B and C is fixed by the
following procedure. Computing the spectrum of elasticity matrix, we extract the eigenvectors,
${\bf e}_1$ and ${\bf e}_2$, corresponding to slowest relaxation modes. Then the points A, B and C
are defined by the condition that the projections of the average changes of $|{\rm AB}|$ onto ${\bf
e}_1$ and of $|{\rm BC}|$ onto ${\bf e}_2$ are maximal.

\end{appendix}

\end{document}